\PassOptionsToPackage{table}{xcolor}
\documentclass[11pt]{article}

\usepackage[preprint]{acl}

\usepackage{times}
\usepackage{latexsym}

\usepackage[T1]{fontenc}

\usepackage[utf8]{inputenc}

\usepackage{microtype}

\usepackage{inconsolata}

\usepackage{graphicx}

\usepackage{tabularx}

\usepackage{booktabs}
\usepackage{multirow}
\usepackage{makecell}
\usepackage{tabularx}
\usepackage{amsmath}
\usepackage{pgfplots}
\pgfplotsset{compat=1.14}
\usepackage{tikz}
\usetikzlibrary{patterns}

\usepackage[table]{xcolor}
\usepackage{amssymb}

\definecolor{freshblue}{RGB}{102,169,214}
\definecolor{freshmint}{RGB}{127,193,168}
\definecolor{freshcoral}{RGB}{240,150,120}

\usepackage{algorithm}
\usepackage{algpseudocode}

\usepackage{arydshln}
\definecolor{algblue}{RGB}{40,90,180}

\algrenewcommand{\algorithmiccomment}[1]{%
\hfill{\footnotesize\textcolor{algblue}{// #1}}%
}

\def\ourmethod{{\textsc{DirectAudioEdit}}}
\title{\ourmethod{}: Inversion-Free Text-Guided Audio Editing via Diffusion Prediction Contrast
}

\author{
    Zhengkun Ge\textsuperscript{1}, 
    Xiaoqian Liu\textsuperscript{1}, 
    Haoran Zhang\textsuperscript{1},
    Yuan Ge\textsuperscript{1},
    Junxiang Zhang\textsuperscript{1},\\
    {\bf Zhengtao Yu\textsuperscript{2}}, 
    {\bf Jingbo Zhu\textsuperscript{1,3}},
    {\bf Tong Xiao\textsuperscript{1,3}}\thanks{~~Corresponding author.}\\
    \textsuperscript{1}School of Computer Science and Engineering, Northeastern University, Shenyang, China\\
    \textsuperscript{2}Kunming University of Science and Technology
    \textsuperscript{3}NiuTrans Research, Shenyang, China\\
    \texttt{liuxiaoqian0319@outlook.com}
    \texttt{xiaotong@mail.neu.edu.cn}
}

\begin{document}
\maketitle
\begin{abstract}

Text-guided audio editing aims to modify the language-specified acoustic content while preserving edit-irrelevant source components.
Existing training-free methods typically rely on inversion-based editing.
While inversion-free editing is appealing as it decreases computational overhead and reconstruction errors, it remains largely unexplored for audio editing.
The key challenge is to construct a source-to-target editing path through diffusion denoising dynamics.
In this paper, we introduce \ourmethod{}, the first attempt to develop a training-free and inversion-free method for audio editing.
Experiments on music and event-level benchmarks across two backbones show that \ourmethod{} reduces macro-averaged FAD and KL by 15.9\% and 15.8\% compared with DDPM inversion, while achieving up to 64.5\% editing speedup
\footnote{~~Project: \url{https://directaudioedit.github.io/}}.

\end{abstract}

\section{Introduction}

Audio editing aims to modify an existing audio signal according to a user-specified instruction while preserving the irrelevant content of the source audio \citep{wang2023audit, liang2024wavcraft, lan2025guiding}. For example, given an audio clip and an instruction such as “replace the guitar music with a piano melody”, an audio editing system is expected to selectively alter the target attribute while maintaining the remaining acoustic structure. 
Training-free audio editing has attracted increasing attention, as it seeks to adapt powerful pretrained generative audio models to editing tasks without relying on costly paired supervision \citep{manor2024zero,jia2025audioeditor,liang2025audiomorphix}.

\begin{figure}[t]
    \centering
    \includegraphics[width=0.9\linewidth]{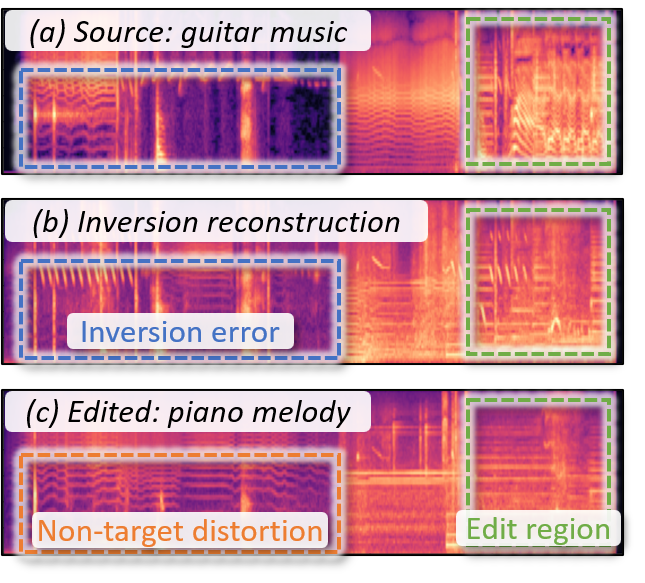}
\caption{
Mel-spectrograms of (a) the source audio, (b) the DDIM-inversion reconstruction, and (c) the edited audio produced by Tango2.
The highlighted regions show that inversion-induced deviations can be carried into target-conditioned editing, causing distortion beyond the intended guitar-to-piano edit.
}

    \label{fig:motivation}
\end{figure}

Beyond audio editing, condition-guided content editing spans a wide range of modalities, including images, audio, and 3D assets \citep{shuai2024survey, huang2025diffusion, lu2024advances}. Across these modalities, editing methods can be broadly summarized into three paradigms: noise-injection-based editing, inversion-based editing, and inversion-free editing. 
Noise-injection-based editing directly injects random noise into the edited region of the source content and then performs condition-guided denoising to synthesize the target content \citep{meng2021sdedit}. 
Inversion-based editing first obtains an initial noise through image-to-noise inversion of the source content and then applies a similar condition-guided denoising process from the inverted noise toward the edited output \citep{10205188, huberman2024edit}.
Inversion-free editing directly transforms the source content under instruction guidance without explicitly reconstructing its generative path \citep{kulikov2025flowedit}. 
Among them, inversion-based audio reconstruction and editing causes undesired distortion, as shown in Fig \ref{fig:motivation}. 
In contrast, inversion-free editing is particularly appealing as it decreases the computational overhead and reconstruction errors \citep{kim2025flowalign, dao2026steerflow}. 
However, inversion-free method remains unexplored for audio editing.

In this paper, we investigate inversion-free audio editing with pretrained text-to-audio generative models. A key challenge is that mainstream audio generative models are commonly built upon diffusion models rather than flow models \citep{majumder2024tango, liu2024audioldm}. We find that directly applying inversion-free editing, such as \textsc{FlowEdit} \citep{kulikov2025flowedit}, to diffusion-based audio models is suboptimal. The target editing path is harder to estimate because the generation paths of diffusion models described by stochastic differential equations are curved and uncertain.

To this end, we propose \ourmethod{}, a training-free and inversion-free method for diffusion-based audio editing.
Rather than indirectly estimating the target state through path conversion, \ourmethod{} directly constructs a more accurate target state estimate by conditioning on the initial noise available during both inversion and denoising.
A dynamic guidance schedule further balances editing strength and source preservation.
Experiments on music and event-level editing benchmarks across two diffusion backbones show that \ourmethod{} reduces macro-averaged FAD and KL by 15.9\% and 15.8\% over DDPM inversion with comparable target alignment. 
We summarize our contributions as follows:
\begin{itemize}
    \item We make the first attempt to extend inversion-free editing to the challenging audio editing task, aiming to improve both editing efficiency and editing quality without relying on costly inversion procedures.

    \item We propose \ourmethod{}, an inversion-free audio editing method that improves target state estimation by leveraging the shared initial noise available throughout both the inversion and denoising processes.

    \item Extensive experiments on AudioLDM2 and Tango2 demonstrate the effectiveness of \ourmethod{} across music and event-level editing tasks. \ourmethod{} achieves up to 64.5\% acceleration and 85.0\% performance improvement compared with inversion-based editing methods.
\end{itemize}

\section{Preliminary}

\subsection{Content Editing}
Suppose we are given a source content $X^{\text{src}}$, a source prompt $c^{\text{src}}$ describing its original semantics, and a target prompt $c^{\text{tgt}}$ specifying the desired edit, text-guided editing aims to produce
\[
X^{\text{tgt}} = \mathcal{M}_\theta(X^{\text{src}}, c^{\text{src}}, c^{\text{tgt}}),
\]
where $ \mathcal{M}_\theta$ is a pretrained diffusion or flow model, and $X^{\text{tgt}}$ should preserve edit-irrelevant information in $X^{\text{src}}$ while reflecting the target semantics in $c^{\text{tgt}}$. 
For example, in audio editing, a guitar performance can be edited into a piano performance by setting $c^{\text{src}}$ to ``guitar'' and $c^{\text{tgt}}$ to ``piano'', while preserving the original melody.

\subsection{Inversion-based Editing}
\begin{figure*}[t]
    \centering
    \includegraphics[width=0.98\linewidth]{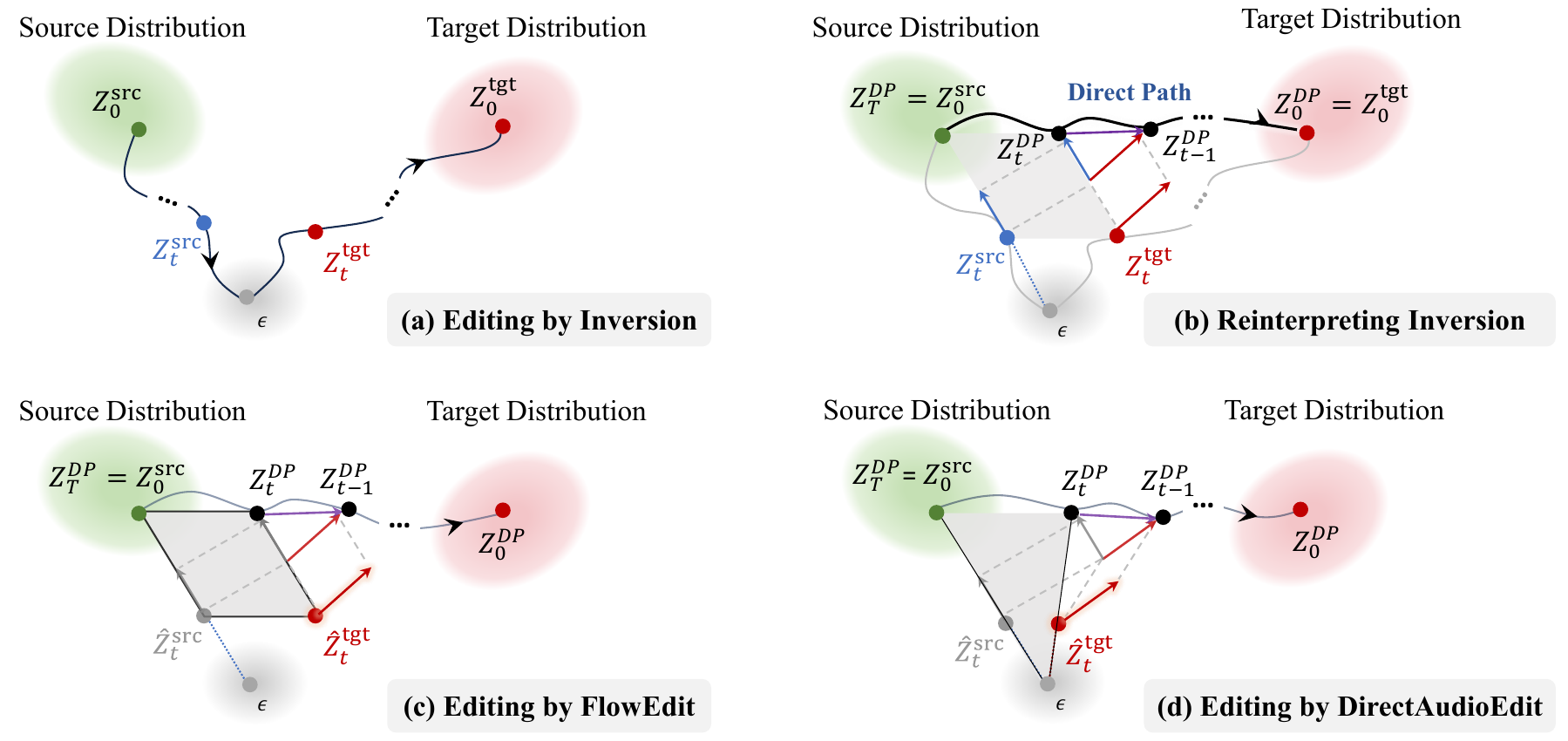}
\caption{
Comparison of editing paradigms.
(a) Inversion-based editing reconstructs a noisy source trajectory before target-guided denoising, while (b) reinterprets it as a direct source-to-target trajectory.
(c) FlowEdit implements inversion-free editing in flow models via flow-direction contrast, whereas (d) DirectAudioEdit implements it in diffusion models via shared-noise re-noising and reverse-dynamics contrast.
}

    \label{fig:main_framework}
\end{figure*}

As shown in Fig. \ref{fig:main_framework}a, inversion-based methods including two process: inversion and denoising. 
The inversion process maps the source content $X^{\text{src}}$ to the initial noise $\epsilon$ under the source condition $c^{\text{src}}$, whereas the denoising process generates the final content $X^{\text{tgt}}$ from the inverted noise $\epsilon$ under the target condition $c^{\text{tgt}}$. 
Specifically, the inversion and denoising paths are respectively defined by the following ordinary differential equations (ODEs):
\begin{align}
    dZ^{\text{src}}_t = V^{\text{src}}(Z^{\text{src}}_t, t)dt\\
    dZ^{\text{tgt}}_t = V^{\text{tgt}}(Z^{\text{tgt}}_t, t)dt
\end{align}
where $V^{\text{src}}$ and $V^{\text{tgt}}$ denote the \textit{text-conditioned velocities} induced by the pre-trained diffusion/flow model under the source and target conditions, $c^{\text{src}}$ and $c^{\text{tgt}}$, respectively.

\subsection{Inversion-free Editing with Direct Path}

As shown in Fig. \ref{fig:main_framework}b, \citet{kulikov2025flowedit} proposes a reinterpretation of inversion-based editing. Specifically, given the forward and reverse flow paths, $Z^{\text{src}}$ and $Z^{\text{tgt}}$, we can construct a virtual \textit{direct path} $Z^{\text{DP}}_t$ defined by an ODE in Eq. \ref{eq:velocity}:
\begin{align}
    Z^{\text{DP}}_t =  Z^{\text{src}}_0 + Z^{\text{tgt}}_t -Z^{\text{src}}_t\\
    dZ^{\text{DP}}_t = V^{\Delta}(Z^{\text{src}}_t, Z^{\text{tgt}}_t, t)dt
    \label{eq:velocity}
\end{align}
where $V^\Delta_t =  V^{\text{tgt}}_t - V^{\text{src}}_t$ is the direct velocity.

Motivated by this, inversion-free methods no longer extract the inverted initial noise to construct a conditioned target path. Instead, \textsc{FlowEdit} estimates the target state utilizing \textit{path conversion}: $\hat{Z}^{\text{tgt}}_t =  Z^{\text{DP}}_t + \hat{Z}^{\text{src}}_t - Z^{\text{src}}_t$, where $\hat{Z}^{\text{src}}_t$ is calculated using the conditioned source path.
Then \textsc{FlowEdit} employs the flow model to predict the source velocity $V^{\text{src}}_t$ and the target velocity $V^{\text{tgt}}_t$. The virtual direct velocity at $Z^{\text{DP}}_t$ is calculated as $V^{\text{tgt}}_t-V^{\text{src}}_t$. The final reconstructed target content is obtained after $T$ iterations.

\subsection{Clean-State Editing Path}

We denote by $Z_t^{\text{DP}}$ the current clean estimate along the editing path.
Subscript $t$ indexes the diffusion timestep used to construct the noisy estimates.
For simplicity, we write the selected editing timesteps as $T,\ldots,1$, where $t-1$ denotes the next lower-noise step after $t$ in the schedule.
The clean editing path is initialized from the source latent:
\begin{equation}
Z_T^{\text{DP}} = Z^{\text{src}}_0,
\end{equation}
where $T$ denotes the initial editing timestep.
The path is then progressively updated from $Z_t^{\text{DP}}$ to $Z_0^{\text{DP}}$ as the timestep decreases.

\section{Methodology}

A key challenge of inversion-free audio editing is the mainstream audio generative models are commonly built upon diffusion models rather than flow models. We find that directly applying inversion-free editing to diffusion-based audio models is suboptimal because the target editing path $\hat{Z}^{\text{tgt}}_t$ is harder to predict: unlike rectified flows \citep{liu2023flow}, whose sampling paths are encouraged to be relatively straight, diffusion models describe generation through stochastic differential equations (SDE), leading to more curved and uncertain paths. 

Furthermore, the generated target audio depends on the pre-trained diffusion model $\mathcal{M}_\theta$, the target condition $c^{\text{tgt}}$, and the initial noise $\epsilon$. Inversion-based editing suggests that an ideal target audio can be obtained by denoising an \textit{appropriate initial noise} derived through inversion. 
To improve efficiency, inversion-free editing instead adopts randomly initialized noise. 
Although this random initialization may be biased, we argue that conditioning the denoising process on the initial noise allows the target state to be progressively corrected, leading to ideal performance.
To this end, we propose \ourmethod{} in this section.

\subsection{Shared-Noise Re-noising}

To address the uncertainty and bias introduced by diffusion paths and randomly initialized noise, we establish a shared-noise re-noising scheme that makes the source and target branches comparable before estimating their editing direction.

We use $\mathcal{R}_t(x;\epsilon)$ to denote the standard forward re-noising operation that maps a clean latent $x$ to timestep $t$ with Gaussian noise $\epsilon \sim \mathcal{N}(0,I)$:
\begin{equation}
\mathcal{R}_t(x;\epsilon)
=
\sqrt{\alpha_t}x
+
\sqrt{1-\alpha_t}\epsilon,
\end{equation}
where $\alpha_t$ is the cumulative noise coefficient of the diffusion schedule.

For the source branch, the clean reference is directly available as $Z^{\mathrm{src}}_0$.
Given the shared noise $\epsilon_t$, we construct the source-side noisy estimate as
\begin{equation}
\hat{Z}^{\mathrm{src}}_t
=
\mathcal{R}_t(Z^{\mathrm{src}}_0;\epsilon_t).
\end{equation}
This estimate anchors the source-conditioned reverse dynamics around the original source content.
For the target branch, we use the current clean editing state $Z_t^{\text{DP}}$ to predict the target-side noisy estimate $\hat{Z}^{\mathrm{tgt}}_t$.
Although $Z_t^{\text{DP}}$ is not the true target latent, it provides the current clean estimate along the editing path.
Initialized from the source latent, $Z_t^{\text{DP}}$ progressively accumulates target-oriented updates along the clean-state editing path.
We construct the target-side noisy estimate as
\begin{equation}
\hat{Z}^{\mathrm{tgt}}_t
=
\mathcal{R}_t(Z_t^{\text{DP}};\epsilon_t).
\end{equation}
This estimate evaluates the target-conditioned reverse dynamics around the current edited state, indicating how the clean editing path should further move toward the target condition.

\subsection{Diffusion Prediction Contrast}

With the source-side and target-side noisy estimates constructed, we next compare their local reverse dynamics.
Let $\Phi_\theta$ denote one reverse scheduler update from timestep $t$ to $t-1$ induced by the pretrained diffusion model.
For the source-side estimate, we compute
\begin{equation}
\hat{Z}^{\mathrm{src}}_{t-1}
=
\Phi_\theta
\left(
\hat{Z}^{\mathrm{src}}_t,
t,
c^{\mathrm{src}};
w^{\mathrm{src}}(t)
\right).
\end{equation}
The source-side reverse displacement is defined as
\begin{equation}
\Delta^{\mathrm{src}}_t
=
\hat{Z}^{\mathrm{src}}_{t-1}
-
\hat{Z}^{\mathrm{src}}_t .
\end{equation}
Here, $c^{\mathrm{src}}$ and $w^{\mathrm{src}}(t)$ denote the source text condition and the guidance scale. The target-side prediction $\hat{Z}^{\mathrm{tgt}}_{t-1}$ and displacement $\Delta^{\mathrm{tgt}}_t$ are obtained analogously by replacing the source-side noisy estimate, condition, and guidance scale with their target-side counterparts. $\Delta^{\mathrm{src}}_t$ describes the source-consistent reverse motion around the original content, while $\Delta^{\mathrm{tgt}}_t$ describes the target-conditioned reverse motion around the current edited state.

We define the editing direction as the diffusion prediction contrast between the target- and source-conditioned reverse displacements:
\begin{equation}
\Delta^{\mathrm{edit}}_t
=
\Delta^{\mathrm{tgt}}_t
-
\Delta^{\mathrm{src}}_t .
\end{equation}
This contrast estimates the target-induced motion relative to the source dynamics.
Because the two estimates are constructed with shared noise, $\Delta^{\mathrm{edit}}_t$ mainly reflects the difference between the source- and target-conditioned reverse dynamics. We use this contrast to update the clean editing state:
\begin{equation}
Z_{t-1}^{\text{DP}}
=
Z_t^{\text{DP}}
+
\Delta^{\mathrm{edit}}_t .
\end{equation}
Although $\Delta^{\text{edit}}_t$ is estimated from noisy diffusion estimates, we do not treat it as a reconstructed noisy latent, but as a relative residual direction for clean-state editing. Since the source and target estimates share the same forward kernel, timestep, and noise realization, their displacement difference suppresses shared stochastic effects and highlights the condition-induced deviation between source- and target-conditioned reverse dynamics. \ourmethod{} applies this residual as a first-order correction to $Z_t^{\text{DP}}$, rather than following an inverted noisy path. Repeating this update steers the clean editing path toward the target semantics while retaining source-conditioned structure, yielding the final edited latent $Z_0^{\text{DP}}$.

\subsection{Dynamic Guidance Schedule}
\label{sec:guidance_scheduling}

\begin{algorithm}[t]
\caption{\ourmethod{}}
\label{alg:direct_audio_edit}
{\small
\begin{algorithmic}[1]
\Require $Z^{\mathrm{src}}_0$, $c^{\mathrm{src}}$, $c^{\mathrm{tgt}}$, $\mathcal{R}_t$, $\Phi_\theta$, timesteps $\{T,\ldots,1\}$, branch-wise guidance scales $w^b(t)$
\Ensure Edited latent $Z_0^{\text{DP}}$

\State $Z_t^{\text{DP}} \leftarrow Z^{\mathrm{src}}_0$ \Comment{initialize}

\For{$t = T, T-1, \ldots, 1$}
    \State $\epsilon_t \sim \mathcal{N}(0,I)$ \Comment{shared noise}
    \State $X^{\mathrm{src}} \leftarrow Z^{\mathrm{src}}_0,\quad X^{\mathrm{tgt}} \leftarrow Z_t^{\text{DP}}$ \Comment{clean references}

    \For{$b \in \{\mathrm{src},\mathrm{tgt}\}$}
        \State $\hat{Z}^b_t \leftarrow \mathcal{R}_t(X^b;\epsilon_t)$ \Comment{re-noise}
        \State $\hat{Z}^b_{t-1} \leftarrow \Phi_\theta(\hat{Z}^b_t,t,c^b;w^b(t))$ \Comment{reverse update}
        \State $\Delta^b_t \leftarrow \hat{Z}^b_{t-1}-\hat{Z}^b_t$ \Comment{displacement}
    \EndFor

    \State $\Delta^{\mathrm{edit}}_t \leftarrow \Delta^{\mathrm{tgt}}_t-\Delta^{\mathrm{src}}_t$ \Comment{prediction contrast}
    \State $Z_{t-1}^{\text{DP}} \leftarrow Z_t^{\text{DP}}+\Delta^{\mathrm{edit}}_t$ \Comment{clean-state update}
\EndFor

\State \Return $Z_0^{\text{DP}}$

\end{algorithmic}
}
\end{algorithm}

Classifier-free guidance (CFG) \citep{ho2022classifierfree} modulates how strongly each reverse update follows its text condition.
In \ourmethod{}, the source and target branches serve different roles and therefore use different guidance strategies.
The source branch acts as a stable structural reference for the original audio, so we keep its guidance scale fixed throughout editing:
\begin{equation}
w^{\mathrm{src}}(t)=w^{\mathrm{src}} .
\end{equation}
The target branch is responsible for injecting the desired edit semantics; instead of using static guidance, we progressively strengthen its guidance scale along the clean-state editing path:
\begin{equation}
w^{\mathrm{tgt}}(t)
=
w^{\min}
+
s_t
\left(
w^{\max}-w^{\min}
\right),
\end{equation}
where $s_t$ is a monotonic schedule that increases from $0$ to $1$ as the editing process proceeds from $Z_t^{\text{DP}}$ to $Z_0^{\text{DP}}$.
More details about the concrete form and hyperparameter values of this schedule are provided in Appendix~\ref{app:implementation_details}.

This scheduling reflects the changing role of target guidance during editing.
At early timesteps, the editing state remains close to the source latent, and weaker target guidance helps avoid abrupt deviations from edit-irrelevant source structure.
As editing proceeds, stronger target guidance encourages more complete target-semantic injection.
This asymmetric design keeps the source branch as a fixed reference field while allowing the target branch to control the strength of semantic deviation. The full process is presented in Algorithm~\ref{alg:direct_audio_edit}.

\section{Experiments}

\begin{table*}[t]
\centering
\scriptsize
\setlength{\tabcolsep}{4pt}
\newcommand{\best}[1]{\textbf{#1}}
\newcommand{\second}[1]{\underline{#1}}
\resizebox{\textwidth}{!}{
\begin{tabular}{llcccccccccc}
\toprule
\multirow{2}{*}{\textbf{Task}} & \multirow{2}{*}{\textbf{Method}}
& \multicolumn{5}{c}{\textbf{AudioLDM2}}
& \multicolumn{5}{c}{\textbf{Tango2}} \\
\cmidrule(lr){3-7} \cmidrule(lr){8-12}
& & \textbf{CLAP $\uparrow$} & \textbf{FAD $\downarrow$} & \textbf{KL $\downarrow$} & \textbf{IS $\uparrow$} & \textbf{SSIM $\uparrow$}
  & \textbf{CLAP $\uparrow$} & \textbf{FAD $\downarrow$} & \textbf{KL $\downarrow$} & \textbf{IS $\uparrow$} & \textbf{SSIM $\uparrow$} \\
\midrule

\rowcolor{gray!12}
\multicolumn{12}{c}{\textit{\textbf{Event-level Editing Benchmark}}} \\

\multirow{4}{*}{Addition}
& SDEdit   & \best{43.03} & 2.166 & 1.257 & 6.626 & 0.311 & \best{47.74} & 4.049 & 1.020 & 5.886 & 0.354 \\
& DDIM-Inv & 34.16 & 3.428 & 2.154 & 6.073 & 0.535 & 39.38 & 6.547 & 2.094 & 5.251 & 0.479 \\
& DDPM-Inv & \second{42.23} & \second{1.046} & \second{0.721} & \best{6.857} & \second{0.688} & 47.28 & \second{3.500} & \second{0.796} & \second{6.239} & \second{0.634} \\
& \textbf{\ourmethod{}} & 41.44 & \best{0.989} & \best{0.536} & \second{6.709} & \best{0.702} & \second{47.65} & \best{3.103} & \best{0.674} & \best{6.709} & \best{0.655} \\

\midrule
\multirow{4}{*}{Removal}
& SDEdit   & \second{40.09} & 2.852 & 2.160 & 6.706 & 0.281 & 41.67 & 3.717 & 1.467 & 5.858 & 0.361 \\
& DDIM-Inv & 35.88 & 4.690 & 2.748 & 6.363 & 0.451 & 23.71 & 11.425 & 3.423 & 4.653 & 0.410 \\
& DDPM-Inv & \best{40.56} & \best{1.660} & \second{1.590} & \second{6.793} & \best{0.615} & \second{41.75} & \second{3.395} & \best{1.298} & \second{6.245} & \second{0.647} \\
& \textbf{\ourmethod{}} & 39.18 & \second{2.229} & \best{1.393} & \best{6.848} & \second{0.604} & \best{43.87} & \best{2.923} & \second{1.312} & \best{6.699} & \best{0.658} \\

\midrule
\multirow{4}{*}{Replacement}
& SDEdit   & \best{43.43} & 2.466 & 1.443 & 7.144 & 0.309 & \second{45.16} & 3.328 & 1.072 & 6.170 & 0.357 \\
& DDIM-Inv & 34.84 & 4.251 & 2.556 & 6.563 & 0.501 & 32.71 & 7.193 & 2.551 & 4.968 & 0.447 \\
& DDPM-Inv & \second{42.46} & \second{1.173} & \second{0.825} & \best{7.427} & \second{0.681} & 44.94 & \second{2.923} & \second{0.892} & \second{6.583} & \second{0.639} \\
& \textbf{\ourmethod{}} & 41.60 & \best{1.048} & \best{0.621} & \second{7.158} & \best{0.695} & \best{45.82} & \best{2.455} & \best{0.811} & \best{7.060} & \best{0.659} \\

\midrule
\rowcolor{gray!12}
\multicolumn{12}{c}{\textit{\textbf{Music Editing Benchmark}}} \\

\multirow{4}{*}{Music}
& SDEdit   & \best{40.33} & 5.765 & 1.454 & 1.734 & 0.411 & 29.28 & 7.891 & 1.354 & 1.466 & 0.496 \\
& DDIM-Inv & \second{38.93} & 5.347 & 1.921 & \best{1.990} & 0.544 & \best{34.06} & \second{5.164} & 1.598 & \best{1.786} & 0.580 \\
& DDPM-Inv & 37.21 & \second{2.252} & \second{0.821} & 1.668 & \second{0.770} & 31.13 & 6.158 & \second{0.833} & 1.498 & \second{0.729} \\
& \textbf{\ourmethod{}} & 37.40 & \best{1.929} & \best{0.582} & \second{1.807} & \best{0.788} & \second{33.99} & \best{3.919} & \best{0.617} & \second{1.570} & \best{0.757} \\

\bottomrule
\end{tabular}
}
\caption{
Main results on the event-level and music editing benchmarks.
Scores are averaged over five random seeds.
Best results are in \textbf{bold}, and second-best results are \underline{underlined}.
CLAP scores are multiplied by 100.
Higher CLAP, IS, and SSIM indicate better performance, while lower FAD and KL are better.
Mean--standard deviation results and Tukey's HSD post-hoc tests are reported in Appendix~\ref{app:statistics}.
}
\label{tab:main_results}
\end{table*}

\subsection{Experimental Setup}

\paragraph{Backbones.}
We evaluate \ourmethod{} on two representative diffusion backbones: AudioLDM2~\citep{liu2024audioldm} and Tango2~\citep{majumder2024tango}.
Both models generate audio through iterative diffusion denoising, making them suitable testbeds for examining whether inversion-free audio editing can be formulated directly within diffusion reverse dynamics.
Unless otherwise specified, all methods are evaluated using the same backbone, data split, and metric computation pipeline.

\paragraph{Datasets.}
We conduct experiments on two types of text-guided audio editing benchmarks.

Event-level editing. We construct a benchmark from AudioCaps~\citep{kim2019audiocaps}, covering three fundamental editing operations: addition, removal, and replacement. For each operation, we construct and manually clean 363 source-target text pairs, resulting in 1,089 editing pairs in total. The source and target descriptions differ mainly in the edit-relevant event, so the benchmark evaluates whether a method can introduce localized semantic changes while preserving edit-irrelevant acoustic content. Details of the benchmark construction and examples are provided in Appendix~\ref{app:event_benchmark}.

Music editing. We use the MedleyMDPrompts benchmark~\citep{manor2024zero}.
Following the generation length supported by the evaluated text-to-audio backbones, we use the first 10 seconds of each original music recording.
Compared with event-level editing, music editing places stronger emphasis on preserving global musical structures, including rhythm, timbre, and instrumentation, while still requiring the edited audio to match the target description.

\paragraph{Baselines.}
We use three representative training-free audio editing baselines: SDEdit, DDIM inversion, and DDPM inversion~\citep{manor2024zero}.
SDEdit edits the source by adding noise to an intermediate diffusion state and then performing target-conditioned denoising.
DDIM inversion and DDPM inversion first map the source audio into a noisy latent state or path, and then perform target-conditioned reverse diffusion.
These methods represent the common inversion-based paradigm in training-free diffusion audio editing.
Further implementation details, including editing steps and guidance scales, are provided in Appendix~\ref{app:implementation_details}.

\paragraph{Evaluation metrics.}
We evaluate editing performance from target semantic alignment, audio quality, source preservation, and human perceptual quality.
We report LAION-CLAP target similarity~\citep{laionclap} for target alignment; Fr\'echet Audio Distance (FAD) \cite{DBLP:conf/interspeech/KilgourZRS19}, KL divergence (KL), and Inception Score (IS) \cite{DBLP:conf/nips/SalimansGZCRCC16} for audio quality and distributional consistency; and mel-spectrogram SSIM \cite{wang2004image} for source preservation.
Since text-guided audio editing involves an inherent trade-off between target alignment and source preservation, we analyze these metrics jointly rather than optimizing a single metric in isolation.
We also conduct a complementary Mean Opinion Score (MOS) evaluation on the Tango2 Replacement setting, where 40 listeners rate 20 edited clips per method on a five-point scale with method names hidden.

\paragraph{Statistical analysis.}
Each experiment is repeated with five random seeds, and Table~\ref{tab:main_results} reports the averaged scores.
To assess statistical significance, we conduct Tukey's honestly significant difference (HSD) post-hoc tests \cite{tukey1949comparing} for each task--backbone--metric setting, comparing \ourmethod{} with each baseline.
Significance markers denote $p<0.001$ ($^{***}$), $p<0.01$ ($^{**}$), and $p<0.05$ ($^{*}$).
More mean--standard deviation results and detailed pairwise significance tests are provided in Appendix~\ref{app:statistics}.

\subsection{Main Results}

\paragraph{Overall performance.}
Table~\ref{tab:main_results} reports the results on both event-level and music editing benchmarks.
Across the two backbones and four editing settings, \ourmethod{} generally improves distributional quality and source-structure preservation while maintaining competitive target alignment.
Averaged over all task--backbone settings in Table~\ref{tab:main_results}, \ourmethod{} reduces FAD and KL by 15.9\% and 15.8\%, respectively, compared with DDPM inversion.
Tukey's HSD post-hoc tests over five random seeds further show that the improvements on quality- and preservation-oriented metrics are statistically significant in most comparisons on FAD, KL, and SSIM.
These gains support the feasibility of source-preserving audio editing without explicit inversion.

\paragraph{Event-level editing.}
For event-level editing, Table~\ref{tab:main_results} shows that \ourmethod{} achieves a favorable balance between localized target editing and source preservation across addition, removal, and replacement. The advantage is mainly reflected in distributional quality metrics such as FAD and KL, together with competitive IS and stronger SSIM in most settings. This indicates that diffusion prediction contrast can introduce event-level semantic changes while reducing unintended distortions to the surrounding acoustic scene.

\paragraph{Music editing.}
The music benchmark further stresses source preservation because rhythm, timbre, and instrumentation should remain stable during editing.
As shown in Table~\ref{tab:main_results}, \ourmethod{} achieves the best FAD, KL, and SSIM on both backbones, indicating better preservation of global musical structure.
Although some baselines obtain higher CLAP, their weaker quality or preservation metrics suggest that target alignment alone is insufficient for structure-sensitive audio editing.

\paragraph{Alignment--preservation trade-off.}
The results also show that target-text alignment alone cannot fully reflect editing quality.
Although SDEdit and DDIM inversion achieve higher CLAP in some settings, they often show worse FAD, KL, or SSIM, especially on music editing.
This indicates that the edited audio may better match the target text while losing edit-irrelevant source structures.
By contrast, \ourmethod{} keeps CLAP competitive and achieves stronger quality- and preservation-oriented results.
This supports our motivation: source-preserving audio editing should introduce the target change while avoiding unnecessary distortion to the original audio.

\subsection{Human Evaluation}
We further analyze the MOS results on the Tango2 Replacement setting. Replacement is selected because it requires both suppressing the original source event and introducing the target event, making it representative for perceptual editing assessment. Listeners rate the overall editing quality by jointly considering target-event correctness, source preservation, and audio naturalness. As shown in Table~\ref{tab:mos}, \ourmethod{} obtains the highest average MOS score among compared methods under this setting, indicating that the proposed method produces perceptually competitive edited audio.

\begin{table}[t]
\centering
\small 
\setlength{\tabcolsep}{0pt} 
\renewcommand{\arraystretch}{1.12} 

\begin{tabular*}{\columnwidth}{@{\extracolsep{\fill}}lllc}
\toprule
\textbf{Backbone} & \textbf{Task} & \textbf{Method} & \textbf{MOS} $\uparrow$ \\

\midrule
\multirow{4}{*}{Tango2} & \multirow{4}{*}{Replacement} & SDEdit & 3.04 \\
                         &                             & DDIM-Inv & 2.96 \\
                         &                             & DDPM-Inv & 3.34 \\
                         &                             & \textbf{\ourmethod} & \textbf{3.43} \\ 
\bottomrule
\end{tabular*} 

\caption{
Human evaluation on the Replacement task with the Tango2 backbone.
Higher MOS scores indicate better perceived editing quality.
}
\label{tab:mos}
\end{table}

\subsection{Inference Cost}
\label{sec:inference_cost}

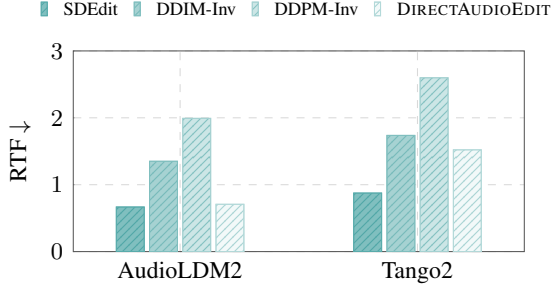
\begin{figure}[t]
\centering
\begin{tikzpicture}
\begin{axis}[
    ybar,
    width=0.98\columnwidth,
    height=0.55\columnwidth,
    bar width=10.5pt,
    ymin=0,
    ymax=3,
    ylabel={RTF $\downarrow$},
    symbolic x coords={AudioLDM2,Tango2},
    xtick=data,
    enlarge x limits=0.45,
    xmajorgrids=true,
    ymajorgrids=true,
    grid style={dashed, gray!30, line width=0.4pt},
    axis line style={black!65},
    tick pos=left,
    tick align=outside,
    major tick length=0pt,
    tick style={black!65},
    tick label style={font=\small},
    label style={font=\small},
    legend style={
        at={(0.5,1.10)},
        anchor=south,
        legend columns=4,
        draw=none,
        font=\scriptsize,
        cells={anchor=west},
        column sep=4pt
    },
    legend image code/.code={
        \draw[#1] (0cm,-0.06cm) rectangle (0.08cm,0.12cm);
    },
]

\addplot[
    draw=teal!65,
    line width=0.1pt,
    fill=teal!50,
    postaction={pattern=north east lines, pattern color=teal!68},
]
coordinates {
    (AudioLDM2,0.6665)
    (Tango2,0.8750)
};
\addlegendentry{SDEdit}

\addplot[
    draw=teal!50,
    line width=0.1pt,
    fill=teal!35,
    postaction={pattern=north east lines, pattern color=teal!53},
]
coordinates {
    (AudioLDM2,1.3503)
    (Tango2,1.7370)
};
\addlegendentry{DDIM-Inv}

\addplot[
    draw=teal!35,
    line width=0.1pt,
    fill=teal!20,
    postaction={pattern=north east lines, pattern color=teal!48},
]
coordinates {
    (AudioLDM2,1.9918)
    (Tango2,2.5984)
};
\addlegendentry{DDPM-Inv}

\addplot[
    draw=teal!30,
    line width=0.1pt,
    fill=teal!5,
    postaction={pattern=north east lines, pattern color=teal!33},
]
coordinates {
    (AudioLDM2,0.7071)
    (Tango2,1.5212)
};
\addlegendentry{\ourmethod{}}

\end{axis}
\end{tikzpicture}
\caption{
RTF comparison of different editing methods on AudioLDM2 and Tango2, measured on a single NVIDIA RTX 3090 GPU.
Lower RTF indicates higher inference efficiency.
}
\label{fig:rtf_analysis}
\end{figure}

We report real-time factor (RTF) on a single NVIDIA RTX 3090 GPU to compare inference cost. As shown in Figure~\ref{fig:rtf_analysis}, \ourmethod{} avoids explicit inversion and therefore achieves lower RTF than DDIM-Inv and DDPM-Inv on both backbones. While SDEdit is faster, its weaker editing quality in Table~\ref{tab:main_results} indicates a less favorable efficiency-quality trade-off.

\section{Ablation Studies and Analysis}

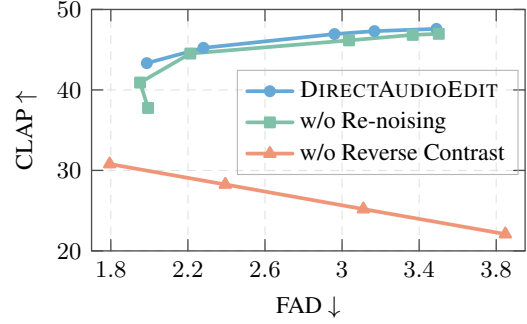
\begin{figure}[t]
\centering
\begin{tikzpicture}
\begin{axis}[
    width=0.95\linewidth,
    height=0.62\linewidth,
    xlabel={FAD $\downarrow$},
    ylabel={CLAP $\uparrow$},
    xmin=1.70, xmax=3.95,
    ymin=20, ymax=50,
    xtick={1.8,2.2,2.6,3.0,3.4,3.8},
    ytick={20,30,40,50},
    grid=major,
    major grid style={dashed, draw=gray!25, line width=0.35pt},
    axis line style={black!70, line width=0.5pt},
    tick style={black!70, line width=0.5pt},
    tick align=inside,
    label style={font=\small},
    tick label style={font=\footnotesize},
    legend style={
        at={(0.66,0.75)},
        anchor=north,
        draw=gray!60,
        fill=none,
        legend columns=1,
        font=\footnotesize,
        /tikz/every even column/.append style={column sep=0.8em}
    },
    legend cell align={left},
    line cap=round,
    line join=round
]

\addplot[
    color=freshblue,
    line width=1.4pt,
    mark=*,
    mark size=1.8pt,
    mark options={draw=freshblue, fill=freshblue, line width=0.9pt}
] coordinates {
    (1.98676,43.3309471)
    (2.28040,45.2284271)
    (2.96065,46.9418833)
    (3.16789,47.3060484)
    (3.48968,47.5866918)

};
\addlegendentry{\ourmethod{}}

\addplot[
    color=freshmint,
    line width=1.4pt,
    mark=square*,
    mark size=1.8pt,
    mark options={draw=freshmint, fill=freshmint, line width=0.9pt}
] coordinates {
    (1.99319,37.7617855)
    (1.95182,40.9129586)
    (2.21232,44.5240786)
    (3.03597,46.1574903)
    (3.36661,46.8386660)
    (3.50201,46.9693704)

};
\addlegendentry{w/o Re-noising}

\addplot[
    color=freshcoral,
    line width=1.4pt,
    mark=triangle*,
    mark size=2.4pt,
    mark options={draw=freshcoral, fill=freshcoral, line width=0.9pt}
] coordinates {
    (1.79379,30.7913487)
    (2.39419,28.2510349)
    (3.11073,25.1914100)
    (3.84758,22.0845773)

};
\addlegendentry{w/o Reverse Contrast}

\end{axis}
\end{tikzpicture}

\caption{CLAP--FAD trade-off for target-state re-noising and reverse-dynamics contrast on the Tango2 Replacement task. Upper-left is better: higher CLAP indicates stronger target alignment, and lower FAD indicates better audio quality.}

\label{fig:ablation_tradeoff}
\end{figure}

\subsection{Why Target-State Re-noising?}

The first ablation examines how the target-side noisy estimate is constructed.
The variant {\color{freshmint}\textbf{w/o Re-noising}} in Figure~\ref{fig:ablation_tradeoff} replaces re-noising $Z_t^{\text{DP}}$ with the translated source-side noisy state defined in Eq.~(3), while keeping the reverse-dynamics update unchanged.
This variant leads to a weaker CLAP--FAD trade-off, indicating that the target branch should be evaluated around the current clean editing state.
Otherwise, the target-side estimate is no longer generated from the clean editing path through the diffusion forward kernel, which can introduce mismatch between the estimated editing direction and the state being updated.
These results show that directly re-noising $Z_t^{\text{DP}}$ is important for obtaining a diffusion-consistent target estimate and maintaining audio quality.

\subsection{Why Reverse-Dynamics Contrast?}

The second ablation studies how the editing direction is computed once the source and target estimates are constructed.
In Figure~\ref{fig:ablation_tradeoff}, {\color{freshcoral}\textbf{w/o Reverse Contrast}} removes the one-step reverse-displacement comparison and instead applies the raw difference between source- and target-conditioned noise predictions.
This variant substantially underperforms the full model, suggesting that prediction differences alone do not provide a reliable update direction for clean-state editing.
By incorporating the scheduler-level reverse update, diffusion prediction contrast accounts for how model predictions are converted into reverse transitions, yielding a more stable local direction for source-to-target editing.

\subsection{Why Dynamic Guidance?}

We further study whether the target-branch guidance should remain fixed during editing.
The variant w/o Dynamic CFG replaces the increasing target CFG schedule with a fixed target guidance scale, while keeping the source guidance and other components unchanged.
As shown in Table~\ref{tab:dynamic_cfg}, dynamic guidance consistently improves FAD, KL, and SSIM on both backbones, indicating better audio quality and source preservation.
On AudioLDM2, it also improves CLAP and IS, whereas on Tango2, CLAP and IS slightly decrease. This discrepancy suggests that dynamic CFG's effect depends on how each backbone converts stronger text guidance into semantic change versus acoustic distortion.
Overall, dynamic guidance acts as a stepwise trade-off controller: weaker target guidance at early steps avoids abrupt source distortion, while stronger guidance at later steps gradually promotes target-semantic injection.

\begin{table}[t]
\centering
\scriptsize
\setlength{\tabcolsep}{2.5pt}
\renewcommand{\arraystretch}{0.92}
\resizebox{\columnwidth}{!}{
\begin{tabular}{lccccc}
\toprule
\textbf{Method} & \textbf{CLAP} $\uparrow$ & \textbf{FAD} $\downarrow$ & \textbf{KL} $\downarrow$ & \textbf{IS} $\uparrow$ & \textbf{SSIM} $\uparrow$ \\
\midrule
\rowcolor{gray!10}
\multicolumn{6}{c}{\textbf{\textit{Backbone: AudioLDM2}}} \\
\rule{0pt}{2ex}w/o Dynamic CFG & 0.410 & 1.184 & 0.658 & 7.207 & 0.689 \\
w Dynamic CFG  & \textbf{0.416} & \textbf{1.099} & \textbf{0.636} & \textbf{7.244} & \textbf{0.695} \\
\midrule
\rowcolor{gray!10}
\multicolumn{6}{c}{\textbf{\textit{Backbone: Tango2}}} \\
\rule{0pt}{2ex}w/o Dynamic CFG & \textbf{0.463} & 2.629 & 0.867 & \textbf{7.147} & 0.647 \\
w Dynamic CFG  & 0.456 & \textbf{2.535} & \textbf{0.821} & 7.039 & \textbf{0.658} \\
\bottomrule
\end{tabular}
}
\caption{
Effect of dynamic target guidance on the Replacement task. The w/o Dynamic CFG setting replaces the increasing CFG schedule with a fixed scale.
}
\label{tab:dynamic_cfg}
\end{table}

\subsection{Qualitative Analysis}

\begin{figure}[t]
    \centering
    \includegraphics[width=0.8\linewidth]{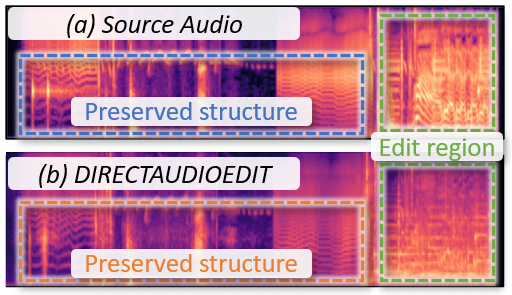}
\caption{
Qualitative case study of \ourmethod{}.
\ourmethod{} introduces the target edit while preserving the main source structure.
}
\label{fig:case_study}
\end{figure}

We further provide a qualitative case study to visualize how \ourmethod{} edits the source audio.
As shown in Figure~\ref{fig:case_study}, the edit-relevant region is modified to introduce the target semantics, while the main spectrogram structure remains largely preserved.
This suggests that \ourmethod{} does not simply regenerate audio from the target prompt, but performs a localized source-to-target update.
The preserved edit-irrelevant patterns and the newly introduced target region are consistent with the design of diffusion prediction contrast, which estimates target-induced changes relative to source-conditioned reverse dynamics.

\section{Related Work}

\subsection{Text-guided Audio Editing}
Early text-guided audio editing methods mainly formulate editing as an instruction-following generation problem, where the model learns to transform a source audio according to a textual instruction.
AUDIT \citep{wang2023audit} constructs a multi-task editing framework from source-instruction-target triplets.
For music editing, InstructME \citep{han2024instructme} introduces a diffusion-based framework with multi-scale feature fusion and musical constraints. These training-based editors can learn explicit editing behaviors, but their scalability is limited by the need for paired editing data \citep{huang2024instructspeech}.

\subsection{Training-free Audio Editing}
Text-conditioned audio diffusion models provide generative priors learned from large-scale audio-text data \citep{liu2024audioldm,huang2023make}.
Recent training-free methods adapt such priors to editing by modifying the sampling process rather than retraining the model \citep{niu2026steermusic}.
A common strategy is to first connect the clean source audio to a source-compatible noisy latent state or path, and then perform target-conditioned denoising~\citep{manor2024zero, jia2025audioeditor}.

\subsection{Inversion-free Generative Editing}
Recent inversion-free editing methods suggest that generative editing need not follow the invert-then-generate paradigm.
FlowEdit \citep{kulikov2025flowedit}, for example, performs text-guided editing through direct source-to-target evolution.
This formulation is closely tied to flow-matching models, which learn continuous velocity fields for transporting samples between distributions \citep{esser2024scaling,DBLP:conf/iclr/LipmanCBNL23}.
Under such models, editing can be formulated as source-to-target transport.

\section{Conclusion}

We presented \ourmethod{}, a training-free and inversion-free method for text-guided audio editing with pretrained diffusion models.
\ourmethod{} formulates editing as direct source-to-target evolution within diffusion denoising dynamics.
It realizes this formulation through diffusion prediction contrast, which constructs comparable source/target estimates via shared-noise re-noising, extracts target-induced updates by reverse-dynamics contrast, and controls their strength with dynamic guidance.

\section*{Limitations}

This work focuses on training-free editing with pretrained diffusion backbones. Its editing ability is therefore bounded by the text-following ability, acoustic coverage, and generation quality of the underlying model. In addition, the current evaluation mainly covers event-level and music editing, while more fine-grained temporal, spatial, and compositional editing instructions remain to be explored.

\bibliography{custom}

\appendix

\section*{Appendix}

\section{Implementation Details}
\label{app:implementation_details}

For a fair comparison, we use 50 editing-stage iterations for all methods, while the operation at each iteration depends on the method.
For SDEdit, we initialize the editing path from the midpoint of a 100-step diffusion schedule by adding noise to the source latent according to the corresponding noise level, and then generate the edited result with 50 reverse denoising steps.
DDIM-Inv performs 50 inversion steps and DDPM-Inv performs 100 inversion steps, after which both methods generate the edited result with 50 reverse denoising steps.
\ourmethod{} instead performs 50 updates on the editing state $Z_t^{\text{DP}}$.

For guidance settings, we use a fixed source CFG scale $w_{\text{src}}=3$ whenever source-conditioned guidance is involved. 
For target guidance in the baselines, we follow the recommended CFG settings in~\cite{manor2024zero}, setting the target CFG scales of SDEdit, DDIM-Inv, and DDPM-Inv to 12, 5, and 12, respectively. For \ourmethod{}, the target CFG scale $w_{\text{tgt}}$ follows a cosine schedule from 8 to 12, which is kept fixed across all settings without task-specific tuning.

For evaluation, we compute FAD, KL, IS, and SSIM using the public AudioLDM evaluation toolkit\footnote{\url{https://github.com/haoheliu/audioldm_eval}}~\cite{liu2023audioldm,liu2024audioldm}. 
For semantic alignment, we compute LAION-CLAP cosine similarity~\cite{laionclap} between each generated audio and the target prompt.

\section{Event-level Editing Benchmark}
\label{app:event_benchmark}

We construct the event-level editing benchmark from AudioCaps to evaluate localized text-guided audio editing. Each example consists of a source audio, a source caption, and a target caption. The benchmark covers three edit operations: addition, removal, and replacement. Addition introduces a new event into the source scene, removal suppresses an existing event, and replacement substitutes one event with another while keeping the remaining scene unchanged.

For each selected AudioCaps caption, we prompt a large language model~\cite{deepseekai2024deepseekv3technicalreport} to generate target captions corresponding to the three edit operations. The generated target captions are required to preserve the edit-irrelevant context of the source caption and modify only the edit-relevant acoustic event. We then manually clean the generated pairs by removing ambiguous edits, unrealistic target events, overly broad event descriptions, and cases where the intended edit cannot be clearly localized from the caption. This yields 363 pairs for each operation and 1,089 pairs in total.

Table~\ref{tab:event_examples} provides representative examples.

\begin{table}[t]
\centering
\small
\setlength{\tabcolsep}{4pt}
\renewcommand{\arraystretch}{1.15}

\newcommand{\added}[1]{\ttfamily{\textcolor{green!50!black}{#1}}}
\newcommand{\deleted}[1]{\ttfamily{\textcolor{red}{#1}}}

\begin{tabularx}{\columnwidth}{lX}
\toprule
\textbf{Operation} & \textbf{Description} \\
\midrule

\multirow{2}{*}{Addition}
& \textbf{Source:} A child talking then a man speaking with bird sounds in the background. \\
& \textbf{Target:} A child talking then a man speaking with bird sounds {\added{and distant thunder}} in the background. \\

\midrule
\multirow{2}{*}{Removal}
& \textbf{Source:} A child talking then a man speaking {\deleted{with bird sounds in the background}}. \\
& \textbf{Target:} A child talking then a man speaking. \\

\midrule
\multirow{2}{*}{Replacement}
& \textbf{Source:} A child talking then a {\deleted{man}} speaking with bird sounds in the background. \\
& \textbf{Target:} A child talking then a {\added{woman}} speaking with bird sounds in the background. \\

\bottomrule
\end{tabularx}
\caption{Representative source-target text pairs in the event-level editing benchmark.}
\label{tab:event_examples}
\end{table}

\section{Statistical Results}
\label{app:statistics}

\providecommand{\sigpos}[1]{{\scriptsize +#1}}
\providecommand{\signeg}[1]{{\scriptsize -#1}}
\providecommand{\notsig}{{\scriptsize --}}

Table~\ref{tab:merged_benchmarks} reports the mean and standard deviation of the automatic metrics for both benchmarks.
To assess whether the differences between \ourmethod{} and the baselines are statistically significant, we conduct Tukey's honestly significant difference (HSD) post-hoc tests for each task--backbone--metric setting.
Table~\ref{tab:sig_merged} reports the pairwise significance results.

In the significance tables, \sigpos{***}, \sigpos{**}, and \sigpos{*} indicate that \ourmethod{} is significantly better than the corresponding baseline at $p<0.001$, $p<0.01$, and $p<0.05$, respectively.
Similarly, \signeg{***}, \signeg{**}, and \signeg{*} indicate that \ourmethod{} is significantly worse.
The symbol \notsig{} indicates no significant difference.

\begin{table*}[t]
\centering
\vspace*{-5cm}

\scriptsize
\setlength{\tabcolsep}{3pt}
\resizebox{\textwidth}{!}{
\begin{tabular}{lcccccccccc}
\toprule
\multirow{2}{*}{\textbf{Method}}
& \multicolumn{5}{c}{\textbf{AudioLDM2}}
& \multicolumn{5}{c}{\textbf{Tango2}} \\
\cmidrule(lr){2-6} \cmidrule(lr){7-11}
& \textbf{CLAP $\uparrow$} & \textbf{FAD $\downarrow$} & \textbf{KL $\downarrow$} & \textbf{IS $\uparrow$} & \textbf{SSIM $\uparrow$}
& \textbf{CLAP $\uparrow$} & \textbf{FAD $\downarrow$} & \textbf{KL $\downarrow$} & \textbf{IS $\uparrow$} & \textbf{SSIM $\uparrow$} \\
\midrule

\rowcolor{gray!20}
\multicolumn{11}{c}{\textbf{Event-level Editing Benchmark}} \\
\midrule

\multicolumn{11}{c}{\textbf{\textit{Addition}}} \\
\hdashline
\rule{0pt}{2ex}SDEdit & 43.03$_{\pm 0.39}$ & 2.166$_{\pm 0.316}$ & 1.257$_{\pm 0.034}$ & 6.626$_{\pm 0.242}$ & 0.311$_{\pm 0.002}$ & 47.74$_{\pm 0.18}$ & 4.049$_{\pm 0.134}$ & 1.020$_{\pm 0.021}$ & 5.886$_{\pm 0.082}$ & 0.354$_{\pm 0.001}$ \\
DDIM-Inv & 34.16$_{\pm 0.12}$ & 3.428$_{\pm 0.000}$ & 2.154$_{\pm 0.001}$ & 6.073$_{\pm 0.004}$ & 0.535$_{\pm 0.000}$ & 39.38$_{\pm 0.36}$ & 6.547$_{\pm 0.180}$ & 2.094$_{\pm 0.029}$ & 5.251$_{\pm 0.071}$ & 0.479$_{\pm 0.001}$ \\
DDPM-Inv & 42.23$_{\pm 0.12}$ & 1.046$_{\pm 0.020}$ & 0.721$_{\pm 0.006}$ & 6.857$_{\pm 0.071}$ & 0.688$_{\pm 0.000}$ & 47.28$_{\pm 0.16}$ & 3.500$_{\pm 0.072}$ & 0.796$_{\pm 0.013}$ & 6.239$_{\pm 0.044}$ & 0.634$_{\pm 0.000}$ \\
\ourmethod{} & 41.44$_{\pm 0.12}$ & 0.989$_{\pm 0.029}$ & 0.536$_{\pm 0.009}$ & 6.709$_{\pm 0.034}$ & 0.702$_{\pm 0.001}$ & 47.65$_{\pm 0.12}$ & 3.103$_{\pm 0.083}$ & 0.674$_{\pm 0.005}$ & 6.709$_{\pm 0.051}$ & 0.655$_{\pm 0.001}$ \\

\midrule

\multicolumn{11}{c}{\textbf{\textit{Removal}}} \\
\hdashline
\rule{0pt}{2ex}SDEdit & 40.09$_{\pm 0.36}$ & 2.852$_{\pm 0.216}$ & 2.160$_{\pm 0.056}$ & 6.706$_{\pm 0.237}$ & 0.281$_{\pm 0.002}$ & 41.67$_{\pm 0.26}$ & 3.717$_{\pm 0.098}$ & 1.467$_{\pm 0.042}$ & 5.858$_{\pm 0.081}$ & 0.361$_{\pm 0.001}$ \\
DDIM-Inv & 35.88$_{\pm 0.06}$ & 4.690$_{\pm 0.000}$ & 2.748$_{\pm 0.005}$ & 6.363$_{\pm 0.004}$ & 0.451$_{\pm 0.000}$ & 23.71$_{\pm 0.31}$ & 11.425$_{\pm 0.328}$ & 3.423$_{\pm 0.021}$ & 4.653$_{\pm 0.054}$ & 0.410$_{\pm 0.001}$ \\
DDPM-Inv & 40.56$_{\pm 0.12}$ & 1.660$_{\pm 0.039}$ & 1.590$_{\pm 0.015}$ & 6.793$_{\pm 0.110}$ & 0.615$_{\pm 0.000}$ & 41.75$_{\pm 0.14}$ & 3.395$_{\pm 0.043}$ & 1.298$_{\pm 0.012}$ & 6.245$_{\pm 0.054}$ & 0.647$_{\pm 0.000}$ \\
\ourmethod{} & 39.18$_{\pm 0.12}$ & 2.229$_{\pm 0.014}$ & 1.393$_{\pm 0.011}$ & 6.848$_{\pm 0.039}$ & 0.604$_{\pm 0.001}$ & 43.87$_{\pm 0.06}$ & 2.923$_{\pm 0.052}$ & 1.312$_{\pm 0.011}$ & 6.699$_{\pm 0.035}$ & 0.658$_{\pm 0.001}$ \\

\midrule

\multicolumn{11}{c}{\textbf{\textit{Replacement}}} \\
\hdashline
\rule{0pt}{2ex}SDEdit & 43.43$_{\pm 0.48}$ & 2.466$_{\pm 0.242}$ & 1.443$_{\pm 0.029}$ & 7.144$_{\pm 0.169}$ & 0.309$_{\pm 0.002}$ & 45.16$_{\pm 0.20}$ & 3.328$_{\pm 0.149}$ & 1.072$_{\pm 0.003}$ & 6.170$_{\pm 0.113}$ & 0.357$_{\pm 0.001}$ \\
DDIM-Inv & 34.84$_{\pm 0.20}$ & 4.251$_{\pm 0.001}$ & 2.556$_{\pm 0.002}$ & 6.563$_{\pm 0.002}$ & 0.501$_{\pm 0.000}$ & 32.71$_{\pm 0.37}$ & 7.193$_{\pm 0.204}$ & 2.551$_{\pm 0.027}$ & 4.968$_{\pm 0.044}$ & 0.447$_{\pm 0.002}$ \\
DDPM-Inv & 42.46$_{\pm 0.27}$ & 1.173$_{\pm 0.034}$ & 0.825$_{\pm 0.007}$ & 7.427$_{\pm 0.086}$ & 0.681$_{\pm 0.000}$ & 44.94$_{\pm 0.17}$ & 2.923$_{\pm 0.031}$ & 0.892$_{\pm 0.007}$ & 6.583$_{\pm 0.053}$ & 0.639$_{\pm 0.000}$ \\
\ourmethod{} & 41.60$_{\pm 0.05}$ & 1.048$_{\pm 0.031}$ & 0.621$_{\pm 0.021}$ & 7.158$_{\pm 0.071}$ & 0.695$_{\pm 0.001}$ & 45.82$_{\pm 0.15}$ & 2.455$_{\pm 0.053}$ & 0.811$_{\pm 0.014}$ & 7.060$_{\pm 0.020}$ & 0.659$_{\pm 0.001}$ \\

\midrule
\rowcolor{gray!20}
\multicolumn{11}{c}{\textbf{Music Editing Benchmark}} \\
\midrule

\rule{0pt}{2ex}SDEdit & 40.33$_{\pm 0.52}$ & 5.765$_{\pm 0.592}$ & 1.454$_{\pm 0.015}$ & 1.734$_{\pm 0.051}$ & 0.411$_{\pm 0.005}$ & 29.28$_{\pm 0.39}$ & 7.891$_{\pm 0.389}$ & 1.354$_{\pm 0.046}$ & 1.466$_{\pm 0.027}$ & 0.496$_{\pm 0.003}$ \\
DDIM-Inv & 38.93$_{\pm 0.35}$ & 5.347$_{\pm 0.001}$ & 1.921$_{\pm 0.001}$ & 1.990$_{\pm 0.006}$ & 0.544$_{\pm 0.000}$ & 34.06$_{\pm 0.11}$ & 5.164$_{\pm 0.027}$ & 1.598$_{\pm 0.013}$ & 1.786$_{\pm 0.024}$ & 0.580$_{\pm 0.001}$ \\
DDPM-Inv & 37.21$_{\pm 0.45}$ & 2.252$_{\pm 0.051}$ & 0.821$_{\pm 0.012}$ & 1.668$_{\pm 0.015}$ & 0.770$_{\pm 0.000}$ & 31.13$_{\pm 0.15}$ & 6.158$_{\pm 0.035}$ & 0.833$_{\pm 0.006}$ & 1.498$_{\pm 0.006}$ & 0.729$_{\pm 0.000}$ \\
\ourmethod{} & 37.40$_{\pm 0.13}$ & 1.929$_{\pm 0.020}$ & 0.582$_{\pm 0.013}$ & 1.807$_{\pm 0.035}$ & 0.788$_{\pm 0.001}$ & 33.99$_{\pm 0.10}$ & 3.919$_{\pm 0.070}$ & 0.617$_{\pm 0.011}$ & 1.570$_{\pm 0.011}$ & 0.757$_{\pm 0.001}$ \\

\bottomrule
\end{tabular}
}
\caption{
Mean and standard deviation results on both the event-level editing benchmark and the music editing benchmark over five random seeds. CLAP scores are multiplied by 100.
}
\label{tab:merged_benchmarks}

\vspace{1cm}

\scriptsize 
\setlength{\tabcolsep}{4pt} 
\begin{tabular}{lcccccccccc}
\toprule
\multirow{2}{*}{\textbf{Baseline}}
& \multicolumn{5}{c}{\textbf{AudioLDM2}}
& \multicolumn{5}{c}{\textbf{Tango2}} \\
\cmidrule(lr){2-6} \cmidrule(lr){7-11}
& \textbf{CLAP $\uparrow$} & \textbf{FAD $\downarrow$} & \textbf{KL $\downarrow$} & \textbf{IS $\uparrow$} & \textbf{SSIM $\uparrow$}
& \textbf{CLAP $\uparrow$} & \textbf{FAD $\downarrow$} & \textbf{KL $\downarrow$} & \textbf{IS $\uparrow$} & \textbf{SSIM $\uparrow$} \\
\midrule

\rowcolor{gray!20}
\multicolumn{11}{c}{\textbf{Event-level Editing Benchmark}} \\
\midrule

\multicolumn{11}{c}{\textbf{\textit{Addition}}} \\
\hdashline
\rule{0pt}{2ex}SDEdit & \signeg{***} & \sigpos{***} & \sigpos{***} & \notsig & \sigpos{***}
& \notsig & \sigpos{***} & \sigpos{***} & \sigpos{***} & \sigpos{***} \\
DDIM-Inv & \sigpos{***} & \sigpos{***} & \sigpos{***} & \sigpos{***} & \sigpos{***}
& \sigpos{***} & \sigpos{***} & \sigpos{***} & \sigpos{***} & \sigpos{***} \\
DDPM-Inv & \signeg{***} & \notsig & \sigpos{***} & \notsig & \sigpos{***}
& \notsig & \sigpos{***} & \sigpos{***} & \sigpos{***} & \sigpos{***} \\

\midrule
\multicolumn{11}{c}{\textbf{\textit{Removal}}} \\
\hdashline
\rule{0pt}{2ex}SDEdit & \signeg{***} & \sigpos{***} & \sigpos{***} & \notsig & \sigpos{***}
& \sigpos{***} & \sigpos{***} & \sigpos{***} & \sigpos{***} & \sigpos{***} \\
DDIM-Inv & \sigpos{***} & \sigpos{***} & \sigpos{***} & \sigpos{***} & \sigpos{***}
& \sigpos{***} & \sigpos{***} & \sigpos{***} & \sigpos{***} & \sigpos{***} \\
DDPM-Inv & \signeg{***} & \signeg{***} & \sigpos{***} & \notsig & \signeg{***}
& \sigpos{***} & \sigpos{**} & \notsig & \sigpos{***} & \sigpos{***} \\

\midrule
\multicolumn{11}{c}{\textbf{\textit{Replacement}}} \\
\hdashline
\rule{0pt}{2ex}SDEdit & \signeg{***} & \sigpos{***} & \sigpos{***} & \notsig & \sigpos{***}
& \sigpos{**} & \sigpos{***} & \sigpos{***} & \sigpos{***} & \sigpos{***} \\
DDIM-Inv & \sigpos{***} & \sigpos{***} & \sigpos{***} & \sigpos{***} & \sigpos{***}
& \sigpos{***} & \sigpos{***} & \sigpos{***} & \sigpos{***} & \sigpos{***} \\
DDPM-Inv & \signeg{**} & \notsig & \sigpos{***} & \signeg{**} & \sigpos{***}
& \sigpos{***} & \sigpos{***} & \sigpos{***} & \sigpos{***} & \sigpos{***} \\

\midrule
\rowcolor{gray!20}
\multicolumn{11}{c}{\textbf{Music Editing Benchmark}} \\
\midrule

\rule{0pt}{2ex}SDEdit & \signeg{***} & \sigpos{***} & \sigpos{***} & \sigpos{*} & \sigpos{***}
& \sigpos{***} & \sigpos{***} & \sigpos{***} & \sigpos{***} & \sigpos{***} \\
DDIM-Inv & \signeg{***} & \sigpos{***} & \sigpos{***} & \signeg{***} & \sigpos{***}
& \notsig & \sigpos{***} & \sigpos{***} & \signeg{***} & \sigpos{***} \\
DDPM-Inv & \notsig & \notsig & \sigpos{***} & \sigpos{***} & \sigpos{***}
& \sigpos{***} & \sigpos{***} & \sigpos{***} & \sigpos{***} & \sigpos{***} \\

\bottomrule
\end{tabular}
\caption{
Pairwise significance results from Tukey's HSD post-hoc tests on both the event-level editing benchmark and the music editing benchmark.
Each entry compares \ourmethod{} against the corresponding baseline over five random seeds.
\sigpos{} indicates that \ourmethod{} is significantly better, \signeg{} indicates significantly worse, and \notsig{} indicates no significant difference.
}
\label{tab:sig_merged}
\end{table*}

\section{Mel-Spectrogram Comparison}
\label{app:mel_spectrogram}

\begin{figure*}[t]
    \centering
    \includegraphics[width=1.0\linewidth]{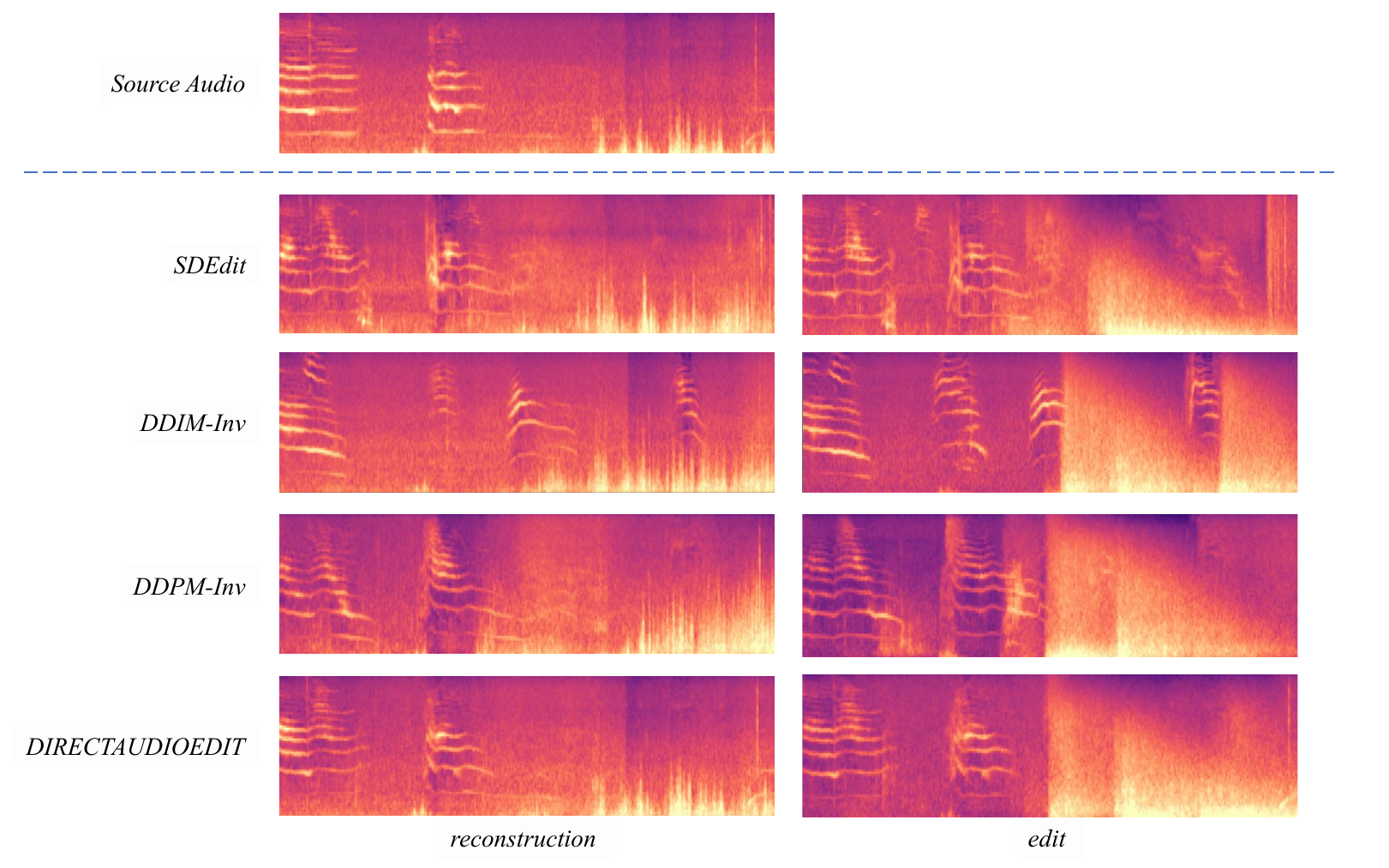}
\caption{
Mel-spectrogram comparison on the Tango2 backbone.
The source prompt is ``Cats meowing and then wind'', and the target prompt is ``Cats meowing and then a violent thunderstorm''.
For each method, the left column shows reconstruction results obtained by setting the target prompt to the source prompt, while the right column shows editing results generated with the target prompt.
}
    \label{fig:mel_comparison}
\end{figure*}

We provide a Mel-spectrogram comparison on the Tango2 backbone in Fig.~\ref{fig:mel_comparison}. The source prompt is ``Cats meowing and then wind'', and the target prompt is ``Cats meowing and then a violent thunderstorm.'' For each method, the left column shows the reconstruction result, where the target prompt is set to the source prompt, and the right column shows the editing result generated with the target prompt.

The visualization illustrates the preservation-editing trade-off of different methods. While baseline methods either introduce noticeable spectral changes or distort part of the source structure after editing, \ourmethod{} better preserves the non-edited cat-meowing region and introduces stronger high-energy patterns in the edited region, which are consistent with the target thunderstorm description. This example suggests that \ourmethod{} can better balance source preservation and target-oriented acoustic modification.

\end{document}